\documentclass[a4paper,11pt]{article}
\pdfoutput=1 

\usepackage{jheppub} 

\usepackage[T1]{fontenc} 

\title{\boldmath Stability and consistent interactions in Podolsky's generalized electrodynamics}

\author[]{J.L. Dai,\note{Corresponding author.}}

\affiliation[a]{Department of Physics, Zhejiang University, Hangzhou, 310027, P. R. China}
\affiliation[b]{Center of Mathematical Science, Zhejiang University, Hangzhou, 310027, P. R. China}

\emailAdd{daijlxy@126.com}

\abstract{We confirm the stability of Podolsky's generalized electrodynamics by constructing a series of two-parametric bounded conserved quantities which includes the canonical energy-momentum tensors. In addition, we evaluate the transition-amplitude of this higher derivative system in BV antifield formalism and obtain the desirable generalized radiation gauge condition by choosing appropriate gauge-fixing fermion. Within the framework of Lagrangian BRST cohomology, we present the constructions of consistent interactions in Podolsky's model and when concentrating on the antighost number zero part of the master action after deformation process, we get the non-Abelian extensions of the Podolsky's theory. Furthermore, we calculate the number of physical degrees of freedom in the resulting higher derivative system utilizing Dirac-Bergmann algorithm method and show that it is unchanged if the consistent interactions are included into the free theory.}

\begin{document}
\maketitle
\flushbottom

\section{Introduction}

The higher order mechanical systems are very active in contemporary physical researchs which have bee analysed from different backgrounds and perspectives ~\cite{1,2,3,4}. As had been surveyed in a series of works, it was realized that Maxwell's $U(1)$ gauge theory is not the unique one to describe the electromagnetic field and in fact the Generalized Electrodynamics, originally proposed by Podolsky was seen to be the most likely appropriate and successful candidate to illustrate the nature of electromagnetic phenomena ~\cite{5,6}. The advantage of this modified form of the Maxwell's electrodynamic theory is that in this formulation, we are able to avoid divergences such as the electron self-energy and the vacuum polarization current which puzzled us for a long time in history ~\cite{7}. Within the framework of Batalin-Fradkin-Vilkovisky (BFV) quantization, it is easy to check that the Podolsky's model has three first-class constraints generating gauge symmetries ~\cite{8,9,10} by means of the canonical Dirac-Bergmann algorithm. Later on, Bufalo and Pimentel ~\cite{11,12} perform a complete BFV analysis of the extensions of the Podolsky's model by adding scalar matter fields and in this way, they acquire the fundamental Green's functions as well as the generalized Ward-Fradkin-Takahashi identities from the generating functional method. On the other hand, motivated by the symplectic structure of the phase space, Nogueira et al. apply the Fadeev-Jackiw symplectic quantization to the reduced order formalism of Podolsky's generalized electrodynamics with the aid of auxiliary fields ~\cite{13}. Indeed, the Fadeev-Jackiw symplectic formalism shows to be more economical compared to the Dirac's quantization scheme which avoids the unnecessary calculations and classification of the constraints in the quantization of gauge systems. Moreover, by the constructions of BRST charge together with BRST-invariant Hamiltonian, Bufalo and Pimentel evaluate the transition-amplitude of this non-Abelian theory by selecting suitable gauge-fixing fermion and they believe that this important outcome is useful for subsequent analysis at the quantum level ~\cite{14}.

However, the theories described by a Lagrangian depending on higher order derivatives with respect to time have quite unsatisfactory property and it is often referred to as the Ostrogradsky instability ~\cite{15,16}. This means that the Hamiltonian of the system is not bounded from below and hence the energy of the dynamics can take an arbitrary negative value. Furthermore, it is impossible to overcome these fatal defects generally by trying to do any alternative canonical transformations. In view of this undesirable and notorious behaviour, recently, Kaparulin, Lyakhovich and Sharapov proposed a new method to handle the issues of stability in higher derivative theories ~\cite{17,18,19,20,21,22}. In their works, a special class of higher derivative models of the so-called derived type were investigated. Concretely, the wave operator determining the equations of motion in these systems can be formulated through a polynomial of arbitrary finite order in another lower order differential operator which is usually named as primary wave operator. In the derived theory, a symmetry of the free higher derivative system means there exists a linear operator that is commutative with the primary wave operator. More importantly, every symmetry of the primary theory will give rise to the $n$-parametric series of higher order symmetries of the field equations if the order of the characteristic polynomial of the wave operator in derived theory is $n$. By performing the Noether's theorem, we are thus led to a series of $n$-parametric conserved quantities ~\cite{18,19}. Especially, the spacetime translation invariance of the action functional can be regarded as the most simplest symmetry of the primary wave operator which will produce a series of conserved second-rank tensors including the standard canonical energy-momentum tensors and the others are different independent integrals of motion. In this way, although the canonical energy is unbounded in higher derivative system, these series of conserved tensors can be bounded by choosing appropriate parameters and thus the theory is stable in classical regime, which also persists at quantum level ~\cite{17}. Inspired by these ideas, the issues of stability in the extended higher Chern-Simons theory coupled to a charged scalar field have been discussed extensively in ~\cite{20,21,22}. Fortunately, due to the wave operator of Podolsky's generalized electrodynamics is factorable, we are able to show that this higher derivative system is also stable both in free and interacting cases. Because of this, they are considered as physically acceptable models.

Since the healthy theories with higher derivative terms possess extra degrees of freedom, these theories are necessarily constrained systems. Moreover, there may exist gauge symmetry in the higher derivative theories analogous to the usual systems which is an essential component for interesting and appealing theories in modern theoretical physics. There is no doubt that the most suitable and powerful tool to deal with the constrained system equipped with gauge symmetry is the BRST formulation developed by Becchi, Rouet, Stora and Tyutin ~\cite{23,24,25}.  Through introducing the anticommutative variables, we can extend the usual gauge symmetry to a larger group namely the BRST symmetry. The BRST transformation is nilpotent and allows us to construct the action in either Lagrangian or Hamiltonian forms, which both of them are equivalent.  Later on, another big prominent advance was made by Henneaux et al, who put forward that the ghost fields that appearing in the BRST transformation rest on an intuitive geometric structure of the Koszul-Tate and longitudinal differentials associated to the constraint surface and gauge orbits ~\cite{26,27,28}. This remarkable fact implies that the homological perturbation theory turns out to be the most appropriate approach to describe the BRST theory ~\cite{29,30} which also permits us to establish the isomorphism between the algebra of observables (the gauge invariant functions) of the original constraint theory and the BRST-cohomology at ghost number zero level ~\cite{31,32,33}.

In this paper, we want to investigate the Batalin-Vilkovisky's consistent interactions in the Podolsky's generalized electrodynamics and this would lead to the non-Abelian generalizations of such higher derivative theory. To begin with, we introduce a collection of $U(1)$ gauge fields to write down the free Lagrangian density with higher derivative terms, from the usual local $U(1)$ gauge symmetries it is simple to construct the extended BV master action in term of the fields and antifields ~\cite{34,35,36}. Then by means of the antibracket, we are able to directly determine the BRST transformation $s$ of this free system which can be decomposed as the sum between the Koszul-Tate differential and the exterior longitudinal derivative along the gauge orbit, namely $s=\delta+\gamma$ ~\cite{29,37}. Subsequently, we deform the extended master action by an expansion of the power series of deformation parameter $g$ and the deformed action $S$ should also satisfy the master equation ~\cite{38,39,40,41,42,43,44,45,46,47,48,49}. In this manner, certainly we receive a series of deformation equations by comparing the order of the parameter $g$ on both sides of the equation and after a detailed analysis, we find that the first-order deformation belongs to the local BRST-cohomology group $H^{0}(s|d)$, here $d$  is the exterior spacetime derivative which will paly an essential role in our derivation. Expounded by the work in ~\cite{31,32}, the evaluation of the elements in $H^{0}(s|d)$ annihilated by $s=\delta+\gamma$ up to a total derivative term can be expanded according to the antighost number. Upon to this result and after a similar discussion, we derive the second-order deformation and the new peculiar feature inherents to the calculation is the emergence of the constraint equations for the coefficients appearing in the first-order deformation action. In fact, the consistency condition satisfied by these coefficients is the famous Jacobi identity and therefore they can be interpreted as the structure constants of some Lie algebra. The other higher order deformation terms can be determined by a straightforward computation following the way we derive the first- and second- order deformation and the procedure will be terminated at the fourth-order deformation ultimately. Finally, adding up all these pieces together, we will gain the desired deformed BV action of the original Lagrangian and the antighost number zero part of this action can be interpreted as non-Abelian version of the generalized electrodynamics in the fundamental representation of some Lie algebra endowed with non-Abelian gauge symmetries.

The organization of this paper is as follows. In section 2, we investigate the issue of stability and calculate the transition-amplitude of Podolsky's generalized electrodynamics with the generalized radiation gauge condition. Section 3 is devoted to the derivation of the first- and higher-order deformations of the master action by means of the local BRST-cohomology in the Batalin-Vilkovisky formalism. In Section 4, we compute the number of physical degrees of freedom and illustrate the stability of the resulting interacting system. The final section of this paper is for discussion and further works.

\section{Generalized Electrodynamics}
\subsection{Degrees of freedom}
Let us consider the Lagrangian density of Podolsky's generalized electrodynamics described by the gauge fields $A_{\mu}$ in (1+3)-dimensional spacetime with metric $g_{\mu\nu}=\mathrm{diag}(1,-1,-1,-1)$ as follows ~\cite{11}
\begin{equation}
\begin{aligned}
\mathcal{L}=-\frac{1}{4}F_{\mu\nu}F^{\mu\nu}+\frac{1}{2m^{2}}\partial_{\mu}F^{\mu\nu}\partial^{\lambda}F_{\lambda\nu}
\end{aligned}
\end{equation}
here the strength field is
$F_{\mu\nu}=\partial_{\mu}A_{\nu}-\partial_{\nu}A_{\mu}$ and we use the metric $g_{\mu\nu}$ raises and lows the indices.

As is well known that all gauge theories are constrained Hamiltonian systems and in view of this, the Dirac analysis ~\cite{51} is regarded as the most powerful tool to investigate the dynamics systematically in these higher derivative constrained systems. Following the spirit of Ostrogradsky's approach ~\cite{15,16}, it is necessary to introduce additional variables to account for the higher derivative nature
\begin{equation}
\begin{aligned}
\Gamma_{\mu}=\partial_{0}A_{\mu}
\end{aligned}
\end{equation}
and the corresponding canonical momenta $(A_{\mu},\Gamma_{\mu})$ are defined explicitly as
\begin{equation}
\begin{aligned}
\pi^{\mu}=\frac{\partial\mathcal{L}}{\partial \dot{A}_{\mu}}-\partial_{\nu}\frac{\partial\mathcal{L}}{\partial(\partial_{\nu} \dot{A}_{\mu})},\quad \quad \phi^{\mu}=\frac{\partial\mathcal{L}}{\partial \dot{\Gamma}_{\mu}}
\end{aligned}
\end{equation}
then a direct calculation shows
\begin{equation}
\begin{aligned}
\pi^{i}&=F^{i0}+\frac{1}{m^{2}}\partial_{i}\partial_{\lambda}F^{\lambda0}-\frac{1}{m^{2}}\partial_{0}\partial_{\lambda}F^{\lambda i},\quad \quad\quad \pi^{0}=\frac{1}{m^{2}}\partial_{i}\partial_{\lambda}F^{\lambda i},\\
\phi^{i}&=\frac{1}{m^{2}}\partial_{\lambda}F^{\lambda i},\quad \quad\quad \quad \quad \quad  \phi^{0}=0
\end{aligned}
\end{equation}
the basic Poisson brackets between them are given by
\begin{equation}
\begin{aligned}
\{A_{\mu}(x),\pi^{\nu}(y)\}=\delta_{\mu}^{\nu}\delta^{3}(x-y),\quad \quad \{\Gamma_{\mu}(x),\phi^{\nu}(y)\}=\delta_{\mu}^{\nu}\delta^{3}(x-y)
\end{aligned}
\end{equation}
in terms of these variables, the canonical Hamiltonian is thus obtained in a standard manner through the Legendre transformation
\begin{equation}
\begin{aligned}
H_{c}=&\int d^{3}x(\pi^{\mu}\dot{A}_{\mu}+\phi^{\mu}\dot{\Gamma}_{\mu}-\mathcal{L})\\
=&\int d^{3}x(\pi_{0}\Gamma^{0}+\pi_{j}\Gamma^{j}+\frac{m^{2}}{2}\phi_{j}\phi^{j}+\phi_{i}\partial^{i}\Gamma_{0}+\phi_{i}\partial_{j}F^{ij}+\frac{1}{4}F_{ij}F^{ij}\\
&-\frac{1}{2}(\Gamma_{j}-\partial_{j}A_{0})^{2}-\frac{1}{2m^{2}}(\partial^{j}\Gamma_{j}-\partial^{j}\partial_{j}A_{0})^{2})
\end{aligned}
\end{equation}

Roughly speaking, even if the Lagrangian of a free system does not include the higher derivative terms, the number of physical degrees of freedom may be altered if adding consistent interactions among the gauge fields. Especially, once the higher derivatives are involved at the free level, this issue becomes more prominent. In the Dirac's analysis of constrained systems, the first main step is to extract the primary constraints from the definition of the momenta and then the primary constraints are required to be conserved by the time evolution under the operation of the Poisson bracket with the enlarged total Hamiltonian. In this way, the identities will generate the secondary constraints and so on. It is simple to obtain the following primary, secondary and tertiary constraints from (2.4) and (2.6)
\begin{equation}
\begin{aligned}
&\Phi_{1}\equiv \phi^{0}\approx 0,\quad \quad \Phi_{2}\equiv \pi^{0}-\partial_{i}\phi^{i}\approx 0,\quad \quad \Phi_{3}=\partial_{i}\pi^{i}\approx 0
\end{aligned}
\end{equation}
one can easily check  $\Phi_{1},\Phi_{2},\Phi_{3}$ have vanishing brackets with all the other constraints, therefore, we assert that there are three first-class constraints in Podolsky's model. It is instructive to count and compare the degrees of freedom of the free system (2.1) before and after the BRST-BV deformations as will be discussed in subsequent sections. Noting that for each first-class constraint, there is one degree of freedom which is not physically important and has to be removed from the theory. Since the total number of canonical variables in the phase space is 16, thus the number of physical degrees of freedom of the original theory is equal to ~\cite{29,52}
\begin{equation}
\begin{aligned}
\mathcal{N}=(16-2\times 3)/2=5
\end{aligned}
\end{equation}
which is larger than the usual Yang-Mills theory due to the higher derivative nature in the system (2.1) as we can expect.

\subsection{Stability}
Generally, as stated above, the higher derivative theories contain more degrees of freedom per dynamical variable as compared to the lower derivative theories, usually these extra degrees of freedom are ghost-like, and their excitation can lower the energy of the system without bounds. This phenomenon also can be seen directly from (2.6) that the canonical Ostrogradski Hamiltonian is a linear function of some momenta and hence it is unbounded from below. When quantum counterpart is considered, the higher derivative term will violate the unitarity of the theory which implies the appearance of associated negative norm states at the quantum level, commonly known as ghost states. In order to remedy this problem, to begin with, let us write down the equations of motion of the gauge fields $A_{\mu}$
\begin{equation}
\begin{aligned}
\square\partial_{\rho}F^{\rho\mu}+m^{2}\partial_{\rho}F^{\rho\mu}=0
\end{aligned}
\end{equation}
then a crucial observation is that this dynamical system is a derived model ~\cite{19}, to be more precise, the primary wave operator $W$ which defines the primary free field theory is given by
\begin{equation}
\begin{aligned}
 W_{\mu\nu}=\delta_{\mu\nu}\square-\partial_{\mu}\partial_{\nu},\quad \quad \quad \quad  W_{\mu\nu}A_{\nu}=0
\end{aligned}
\end{equation}
with the aid of (2.10), the wave operator in (2.9) has the following factorable structure
\begin{equation}
\begin{aligned}
M=W(W+m^{2}),\quad \quad \quad \quad \quad  M_{\mu\nu}A_{\nu}=0
\end{aligned}
\end{equation}
here $M,W$ are understood as matrix differential operators and the order of the characteristic polynomial of the wave operator is two. Furthermore, to probe this factorisation, it is convenient to introduce two new dynamical variables absorbing the higher derivatives of the original fields
\begin{equation}
\begin{aligned}
\eta_{1}=WA,\quad \quad \eta_{2}=(W+m^{2})A
\end{aligned}
\end{equation}
as well as the following action functional
\begin{equation}
\begin{aligned}
&S_{1}\left[\eta_{1},\eta_{2}\right]=\int d^{4}x\left[\eta_{1}(W+m^{2})\eta_{1}+\eta_{2}W\eta_{2}\right]
\end{aligned}
\end{equation}
then the equations of motion of the new fields are simply expressed as
\begin{equation}
\begin{aligned}
\frac{\delta S_{1}}{\delta \eta_{1}}=(W+m^{2})\eta_{1}=0,\quad \quad \quad \frac{\delta S_{1}}{\delta \eta_{2}}=W\eta_{2}=0
\end{aligned}
\end{equation}
and from these expressions we obtain, by upon substitution of (2.12), the dynamic equations of motions for the gauge fields $A_{\mu}$ in (2.9). This apparent fact signifies that the formulae (2.12) establish a one-to-one correspondence between the solutions of systems (2.1) and (2.13). Therefore, we argue that these two systems are equivalent to each other and can be regarded as two different representations of the same theory which are usually named $A$- and $\eta$-representations. From this point of view, we are capable of demonstrating the issues of stability in Podolsky's generalized electrodynamics by constructing a series of two-parametric conserved quantities.

To show this, it is manifest that the fields $\eta_{1}$ and $\eta_{2}$ are independent degrees of freedom in the action functional which is kept invariant under the spacetime translations $x^{\mu}\rightarrow x^{\mu}-\varepsilon^{\mu}$. Under this observation and with the help of Noether's theorem, we are led to the following conserved currents $J^{\mu}(\eta_{i})$ as well as the canonical energy-momentum tensors by the standard rule
\begin{equation}
\begin{aligned}
\partial_{\mu}J^{\mu}(\eta_{i})=-\epsilon^{\mu}\partial_{\mu}\eta_{i\nu}\frac{\delta S_{1}}{\delta\eta_{i\nu}},\quad \quad J^{\mu}(\eta_{i})=\Theta^{\mu}_{\nu}(\eta_{i})\epsilon^{\nu},\quad \quad i=1,2
\end{aligned}
\end{equation}
the primary outcomes in ~\cite{18} show that any derived theory with $n$-th order characteristic polynomial admits $n$-parameter
series of conserved quantities which can be expressed as the linear combinations of $\Theta^{\mu}_{\nu}(\eta_{i})$. In the present case, we have
\begin{equation}
\begin{aligned}
\Theta^{\mu}_{\nu}=\sum_{i=1}^{2}\beta_{i}\Theta^{\mu}_{\nu}(\eta_{i})
\end{aligned}
\end{equation}
after a simple calculation, it is not hard to gain the explicit expressions of $\Theta^{\mu}_{\nu}(\eta_{i})$
\begin{equation}
\begin{aligned}
\Theta^{\mu}_{\nu}(\eta_{1})&=\frac{1}{4}\delta^{\mu}_{\nu}F^{1}_{\rho\lambda}F^{\rho\lambda}_{1}-F^{1}_{\nu\lambda}F^{\mu\lambda}_{1}-\eta_{1\nu}\partial_{\omega}F^{\omega\mu}_{1}-\frac{1}{2}m^{2}\delta^{\mu}_{\nu}\eta_{1\omega}\eta^{\omega}_{1},\\
\Theta^{\mu}_{\nu}(\eta_{2})&=\frac{1}{4}\delta^{\mu}_{\nu}F_{\rho\lambda}^{2}F^{\rho\lambda}_{2}-F^{2}_{\nu\lambda}F^{\mu\lambda}_{2}-\eta_{2\nu}\partial_{\omega}F^{\omega\mu}_{2}
\end{aligned}
\end{equation}
for the sake of clarity, here we use the notation $F^{\rho\lambda}_{i}=\partial^{\rho}\eta_{i}^{\lambda}-\partial^{\lambda}\eta_{i}^{\rho}$.
It is worth remarking that all of the conserved quantities in this series are related to the time translation symmetry by a Lagrange anchor ~\cite{17}. In particular, notice that the canonical energy-momentum density of the original higher derivative system is included into this two-parametric conserved second-rank tensors, though it is always unbounded.  As interpreted in ~\cite{17,18,19}, the component $\Theta_{0}^{0}$ captures the property we are interested in which has the sense of the energy density of the theory. In this manner, the total energy of the higher derivative system is provided by the integral
\begin{equation}
\begin{aligned}
E=\int d^{3}x\Theta_{0}^{0}
\end{aligned}
\end{equation}
more explicitly, when $\mu=\nu=0$, the 00-component in (2.16) can be simplified in the form of
\begin{equation}
\begin{aligned}
\Theta_{0}^{0}=&\beta_{1}(\frac{1}{4}F_{\rho\lambda}^{1}F_{\rho\lambda}^{1}-\eta_{10}\partial_{\omega}F^{\omega0}_{1}-\frac{1}{2}m^{2}\eta_{1\omega}\eta^{\omega}_{1})+\beta_{2}(\frac{1}{4}F_{\rho\lambda}^{2}F_{\rho\lambda}^{2}-\eta_{20}\partial_{\omega}F^{\omega0}_{2})
\end{aligned}
\end{equation}
on the other hand, making using of the equations of motion (2.14), one simply gets
\begin{equation}
\begin{aligned}
\partial_{\omega}F^{\omega0}_{1}+m^{2}\eta_{1}^{0}=0,\quad \quad \quad  \partial_{\omega}F^{\omega0}_{2}=0
\end{aligned}
\end{equation}
afterwards, in the consideration of the metric $g_{\mu\nu}=(1,-1,-1,-1)$, it is helpful to rewrite the expression of the 00-component  as
\begin{equation}
\begin{aligned}
\Theta_{0}^{0}=&\beta_{1}(\frac{1}{4}F_{\rho\lambda}^{1}F_{\rho\lambda}^{1}+\frac{1}{2}m^{2}\eta_{1\omega}\eta_{1\omega})+\frac{1}{4}\beta_{2}F_{\rho\lambda}^{2}F_{\rho\lambda}^{2}
\end{aligned}
\end{equation}
now once this two-parametric summation contains bounded conserved quantities, we are able to infer that the original higher derivative dynamics is stable and this requirement can be achieved by
\begin{equation}
\begin{aligned}
\beta_{1}>0,\quad \quad\quad \beta_{2}>0
\end{aligned}
\end{equation}
under such condition, clearly the 00-component is both bounded and positive, though the canonical energy is unbounded from below.

\subsection{Batalin-Vilkovisky formalism}

The Batalin-Vilkovisky (BV) formalism ~\cite{34,35,36} or named the antifield formalism in the framework of Lagrangian quantization of gauge systems as a natural generalization of the BRST method is equipped with additional antifields as well as a new canonical structure known as the antibracket. Specifically and for simplicity, focusing on the irreducible gauge theories, in the background of BV formalism one needs to introduce a collection of extra fields $\phi^{*}_{i}$ with ghost number $\mathrm{gh}(\phi^{*}_{i})<0$ called antifields on the $n$-dimensional manifold besides the original fields $\phi^{i}$. If compared to the BRST approach, in that case we have a set of fields $\phi^{i}$ with  ghost number $\mathrm{gh}(\phi^{i})\geq0$ only. Now within the BV framework, we denote the extended set of fields by $\phi^{A}=(A_{\mu},\eta^{a},\bar{\eta}^{a},B^{a})$ containing the classical fields $A_{\mu}$, the ghost fields $\eta^{a}$ associated with the gauge invariance, the antighost fields $\bar{\eta}^{a}$ and the Lagrangian multiplies $B^{a}$ for the gauge fixings together with the corresponding antifields $\phi^{*}_{A}=(A^{*\mu},\eta^{*}_{a},\bar{\eta}^{*}_{a},B^{*}_{a})$ with opposite Grassmann parity. In this manner, one obtains a space constituted by the local functionals of the whole fields which is naturally endowed with an odd Poisson bracket $(\quad,\quad)$, called antibracket and hence this resulting field/antifield space acquires an odd phase space structure. Concretely, for arbitrary two local functionals  $F(\phi^{A},\phi^{*}_{A}),G(\phi^{A},\phi^{*}_{A})$, the antibracket is defined by ~\cite{37}
\begin{equation}
\begin{aligned}
(F,G)=\int_{M}(\frac{\delta_{r}F}{\delta\phi^{A}}\frac{\delta_{l}G}{\delta\phi_{A}^{*}}-\frac{\delta_{r}G}{\delta\phi^{A}}\frac{\delta_{l}F}{\delta\phi_{A}^{*}})d^{n}x
\end{aligned}
\end{equation}
here the summation over $A$ is understand and the $l,r$ superscripts on the functional derivatives denote that they are taken from the left or from the right respectively. In such a way, the fields $\phi^{A}$ and antifields $\phi^{*}_{B}$ behave as coordinates and momenta and we can regard them as conjugate variables. To say more, the antibracket satisfies graded commutation, distribution and Jacobi relations as we can imagine, and in particular, the antibracket has ghost number 1.

 In the irreducible gauge theories, the central role in the BV formalism is the master action $S_{0}$ which encodes all of the necessary information about the original gauge theory including gauge transformations, the equations of motion and the Noether's identities. The master action $S_{0}$ is a functional of ghost number $0$ which is in principle completely determined by requiring ~\cite{26}
 \begin{equation}
\begin{aligned}
 (S_{0},S_{0})=0
 \end{aligned}
\end{equation}
and this is the so-called master equation. On the other hand, it is well known that the solution to the master equation exists and the construction of such master action starts with the classical action as its boundary condition while the higher order terms are added by assigning antifields an irreducible generating set of gauge transformations with gauge parameters replaced by ghosts. Generally speaking, a proper minimal solution of the master equation in irreducible gauge theories is given by ~\cite{29}
\begin{equation}
\begin{aligned}
S_{0}(\phi^{A},\phi^{*}_{A})=\int d^{n}x(\mathcal{L}+A_{a}^{*\mu}R_{b}^{\mu a}\eta^{b}+F^{abc}\eta_{a}^{*}\eta^{b}\eta^{c}+\bar{\eta}_{a}^{*}B_{a})
\end{aligned}
\end{equation}
here $R_{b}^{\mu a}$ are the gauge generators, $F^{abc}$ are the structure functions and the "proper" means that $S_{\alpha\beta}=\partial^{r}_{\alpha}\partial^{l}_{\beta}S_{0}$ is a matrix of rank $N$ on the stationary surface constrained by the equations of motion and $N$ is the number of fields $\phi^{A}$ ~\cite{50}. Indeed, the proper condition makes it possible to give the desired gauge-fixed action in the procedure of quantization of the general gauge theory. Also it is worth to remark here that the total solution of the master action is uniquely defined up to anticanonical transformations as the BRST charge does.

At the quantum level, in order to evaluate the path integral of system (2.25) precisely, a gauge condition is essential to remove redundant degrees of freedom and this aim can be accomplished by adding a non-minimal term ~\cite{29}
\begin{equation}
\begin{aligned}
\int d^{4}x\bar{\eta}^{*}\lambda
\end{aligned}
\end{equation}
which makes no influence on the solution of classical master equation, here $\bar{\eta}$ are the antighosts of ghost number minus one, $\lambda$ are the auxiliary field and $\bar{\eta}^{*},\lambda^{*}$ are the corresponding antifields. In this way, for (2.1), one finds that the full solution of classical master equation with this new non-minimal sector included reads
\begin{equation}
\begin{aligned}
S_{0}=\int d^{4}x(\mathcal{L}+A^{*\mu}\partial_{\mu}\eta+\bar{\eta}^{*}\lambda)
\end{aligned}
\end{equation}
to implement the generalized radiation gauge condition, it is necessary to take a gauge-fixing fermion $\Psi$ of the form
\begin{equation}
\begin{aligned}
\Psi=\int d^{3}x\bar{\eta}(\partial^{\mu}A_{\mu}+\frac{\square}{m^{2}}\partial^{\mu}A_{\mu}-\frac{\alpha}{2}\lambda)
\end{aligned}
\end{equation}
here $\alpha$ is an arbitrary constant and notice that the transition-amplitude of the original gauge theory is independent of $\alpha$. Taking advantage of the gauge-fixing fermion, one could reach the gauge-fixed action $S_{\Psi}$ by inserting (2.28) into the solution $S_{0}\left[\phi^{A},\phi^{*}_{A}\right]$ of the master equation through
\begin{equation}
\begin{aligned}
S_{\Psi}=S_{0}\left[\phi^{A},\phi^{*}_{A}=\frac{\delta\Psi}{\delta\phi^{A}}\right]
\end{aligned}
\end{equation}
elimination of the antifields by means of the gauge-fixing fermion, we have
\begin{equation}
\begin{aligned}
\bar{\eta}^{*}=-(\partial^{\mu}A_{\mu}+\frac{\square}{m^{2}}\partial^{\mu}A_{\mu}-\frac{\alpha}{2}\lambda),\quad \quad  \quad A^{\ast\mu}=\partial^{\mu}\bar{\eta}
\end{aligned}
\end{equation}
and the gauge-fixed action (2.29) turns out to be the effective action
\begin{equation}
\begin{aligned}
S_{\Psi}=\int d^{4}x\left[\mathcal{L}-(\partial^{\mu}A_{\mu}+\frac{\square}{m^{2}}\partial^{\mu}A_{\mu}-\frac{\alpha}{2}\lambda)\lambda-\bar{\eta}\square\eta\right]
\end{aligned}
\end{equation}
here $\square $ is the usual D'Alembert operator. Finally, integration over the auxiliary field $\lambda$ in the path integral, it is simple to acquire the  transition-amplitude of Podolsky's theory as follows
 \begin{equation}
\begin{aligned}
Z=\int\left[DA_{\mu}\right]\left[D\eta\right]\left[D\bar{\eta}\right]\mathrm{exp}\frac{i}{\hbar}\int d^{4}x(\mathcal{L}-\frac{1}{2\alpha}\left[(1+\frac{\square}{m^{2}})\partial^{\mu}A_{\mu}\right]^{2}-\bar{\eta}\square\eta)
\end{aligned}
\end{equation}
in the above expression, we exactly obtain the covariant form for the transition-amplitude with generalized radiation gauge condition
\begin{equation}
\begin{aligned}
(1+m^{-2}\square)\partial^{\mu}A_{\mu}=0
\end{aligned}
\end{equation}
and the ghost fields are decoupled from the gauge fields. As a consequence, one can easily check that the path integral of gauge-fixed action of the antifield formalism agrees with the path integral of gauge-fixed action of the Hamiltonian formalism in ~\cite{11,12}. In the latter case, the integration over the momenta leads to a numerical factor which can be absorbed into the measure.

\section{BRST-BV deformations}
\subsection{Deformation equations}
As an application of the deformation theory of the master action, let us consider the following free Lagrangian density by introducing a set of gauge fields $A_{\mu}^{a}$ for $a=1,...,N$
\begin{equation}
\begin{aligned}
\bar{S}_{0}\left[A_{\mu}^{a}\right]=\sum_{a=1}^{N}\int d^{4}x(-\frac{1}{4}F_{\mu\nu}^{a}F^{\mu\nu}_{a}+\partial_{\mu}F^{\mu\nu}_{a}\partial^{\lambda}F_{\lambda\nu}^{a})
\end{aligned}
\end{equation}
here we use the metric $g_{\mu\nu}$ raises and lows the indices such as
\begin{equation}
\begin{aligned}
F_{\mu\nu}^{a}=\partial_{\mu}A_{\nu}^{a}-\partial_{\nu}A_{\mu}^{a},\quad \quad F_{a}^{\mu\nu}=g^{\mu\alpha}g^{\nu\beta}F_{\alpha\beta}^{a},\quad \quad A^{\mu}_{a}=g^{\mu\nu}A_{\nu}^{a}
\end{aligned}
\end{equation}
obviously, the number of physical degrees of freedom in (3.1) is $5N$ and the total Lagrangian density is invariant under the local gauge transformations for gauge fields
\begin{equation}
\begin{aligned}
\Delta_{\varepsilon} A_{\mu}^{a}=\partial_{\mu}\varepsilon^{a}
\end{aligned}
\end{equation}
for arbitrary functions $\varepsilon^{a}(x)$ and with the aid of this symmetry, the minimal solution of the master action of (3.1) is given by
\begin{equation}
\begin{aligned}
S_{0}\left[A^{a}_{\mu},A^{*\mu}_{a},\eta^{a}\right]=\int d^{4}x(-\frac{1}{4}F_{\mu\nu}^{a}F^{\mu\nu}_{a}+\partial_{\mu}F^{\mu\nu}_{a}\partial^{\lambda}F_{\lambda\nu}^{a}+A^{*\mu}_{a}\partial_{\mu}\eta^{a})
\end{aligned}
\end{equation}
here the repeated indices denote summation and the BRST symmetry is canonically generated by the antibracket and master action $S_{0}$ through the action
\begin{equation}
\begin{aligned}
s=(S_{0},\cdot)
\end{aligned}
\end{equation}
as emphasized in ~\cite{26,29}, the BRST generator $s=\delta+\gamma$ of the general irreducible gauge theory can be divided into two parts called Koszul-Tate differential $\delta$ and the exterior longitudinal derivative $\gamma$ which act on the generators of the BRST complex in the following way
\begin{equation}
\begin{aligned}
&\delta A^{a}_{\mu}=0,\quad \gamma A^{a}_{\mu}=\partial_{\mu}\eta^{a},\quad \quad \delta\eta^{a}=0, \quad \gamma\eta^{a}=0,\\
&\delta A^{*\mu}_{a}=\partial_{\nu}F^{\mu\nu}_{a}+2\partial_{\nu}\partial^{\mu}\partial_{i}F_{a}^{i\nu}-2\partial^{j}\partial_{j}\partial_{i}F_{a}^{i\mu},\quad \quad \gamma A^{*\mu}_{a}=0,\\
&\delta\eta^{*}_{a}=-\partial_{\mu}A^{*\mu}_{a},\quad \quad \gamma\eta^{*}_{a}=0
\end{aligned}
\end{equation}
in addition, we introduce the Grassmann parities, antighost, pureghost and ghost numbers of the whole fields $(\phi^{A},\phi^{*}_{A})$ which are useful for the derivation of the deformations of the master action ~\cite{29}
\begin{equation}
\begin{aligned}
&\epsilon(A_{\mu}^{a})=0,\quad \epsilon(A_{a}^{*\mu})=1,\quad \epsilon(\eta^{a})=1,\quad \epsilon(\eta^{*}_{a})=0,\\
&\mathrm{agh}(A_{\mu}^{a})=0,\quad \mathrm{agh}(A_{a}^{*\mu})=1,\quad \mathrm{agh}(\eta^{a})=0,\quad\mathrm{agh}(\eta^{*}_{a})=2,\\
&\mathrm{pgh}(A_{\mu}^{a})=0,\quad \mathrm{pgh}(A_{a}^{*\mu})=0,\quad \mathrm{pgh}(\eta^{a})=1,\quad \mathrm{pgh}(\eta^{*}_{a})=0,\\
&\mathrm{gh}(A_{\mu}^{a})=0,\quad \quad \mathrm{gh}(A_{a}^{*\mu})=-1,\quad \mathrm{gh}(\eta^{a})=1,\quad\quad  \mathrm{gh}(\eta^{*}_{a})=-2\\
\end{aligned}
\end{equation}

As illustrated in ~\cite{38,39,40}, a consistent deformation of the classical action and the corresponding gauge invariance induces a consistent deformation of the master action which meanwhile is preserved by the master equation we mentioned above. Inspired by this, Bizdadea and his collaborates had done extensive works on the constructions of deformations of the master action through the BRST cohomological approach in various contexts of gauge theories which can be consulted in ~\cite{41,42,43,44,45,46,47}. In detail, after the deformation procedure, there will exist additional interacting terms in the action usually named consistent interactions and the deformed gauge transformations are close on-shell in such resulting action. More precisely, if we do the deformation by the introduction of parameter $g$, then express the deformed master action in terms of the parameter as
\begin{equation}
\begin{aligned}
S=S_{0}+gS_{1}+g^{2}S_{2}+......
\end{aligned}
\end{equation}
and we assume the deformation of the master action is consistent, thus it follows from the above assertion that the deformed quantity should still fulfill the master action
\begin{equation}
\begin{aligned}
(S,S)=0
\end{aligned}
\end{equation}
expanding (3.9) and comparing the power series of $g$ order by order, there is no difficulty in obtaining the following deformation equations of the master action
\begin{equation}
\begin{aligned}
&1:(S_{0},S_{0})=0,\\
&g^{1}:2(S_{0},S_{1})=0,\\
&g^{2}:2(S_{0},S_{2})+(S_{1},S_{1})=0,\\
&g^{3}:(S_{0},S_{3})+(S_{1},S_{2})=0,\\
&g^{4}:2(S_{0},S_{4})+2(S_{1},S_{3})+(S_{2},S_{2})=0,\\
&g^{5}:(S_{0},S_{5})+(S_{1},S_{4})+(S_{2},S_{3})=0,\\
&......
\end{aligned}
\end{equation}

\subsection{First-order deformation}
Now let us solve the deformation equations of the deformed master action in the standard procedure of perturbative expansion order by order. In the beginning, we concentrate on the first-order deformation term in (3.10) which generally takes the form of $S_{1}=\int d^{4}x \omega_{1}$ and satisfies the functional equation
\begin{equation}
\begin{aligned}
0=sS_{1}=\int d^{4}x s\omega_{1}
\end{aligned}
\end{equation}
here $\omega_{1}$ is a local functional and we see that the first-order deformation of the master action is $s$-cocycle modulo the total derivative $d$ at ghost number zero. Using the decomposition of $s=\delta+\gamma$, the above $s$-exact equation is equivalent to
\begin{equation}
\begin{aligned}
s\omega_{1}=\delta \omega_{1}+\gamma\omega_{1}=\partial_{\mu}k^{\mu}_{1}
\end{aligned}
\end{equation}
here $k^{\mu}_{1}$ is a local current functional and the functional $\omega_{1} $ is required to fulfill $\mathrm{gh}(\omega_{1} )=\epsilon(\omega_{1} )=0$. To find out the solution of (3.12), let us expand the $\omega_{1} $ according to the antighost number ~\cite{41,42,43}
\begin{equation}
\begin{aligned}
\omega_{1}=\omega_{1}^{(0)} +\omega_{1}^{(1)} +...+\omega_{1}^{(I)}
\end{aligned}
\end{equation}
the antighost number of $\omega_{1}^{(i)}$ is $i$ and it is evident to see from (3.6) and (3.7) that the Koszul-Tate differential $\delta$ lowers the antighost number whereas the exterior longitudinal derivative $\gamma$ keeps the antighost number. In this way, comparing the antighost number on both sides of (3.12), this equation can be decomposed into a series of recursive equations
\begin{equation}
\begin{aligned}
\gamma\omega_{1}^{(I)}=\partial_{\mu}k^{\mu(I)}_{1},\quad \delta\omega_{1}^{(I)}+\gamma\omega_{1}^{(I-1)}=\partial_{\mu}k^{\mu(I-1)}_{1},\quad \delta\omega_{1}^{(i+1)}+\gamma\omega_{1}^{(i)}=\partial_{\mu}k^{\mu(i)}_{1}
\end{aligned}
\end{equation}
up to total derivative terms for $i=0,...,I-2$ and the terms in (3.13) are determined successively from these equations. As explained in ~\cite{41,43}, the highest antighost number $I$ term should be strictly satisfied $\gamma\omega_{1}^{(I)}=0$ which implies the component $\omega_{1}^{(I)}$ of highest antighost number belongs to $H^{I}(\gamma)$. On the other hand, our analysis depends crucially on the fact that $H_{I}(\delta|d)$ vanishes for $I>2$ for Yang-Mills type theory ~\cite{31,32}, and consequently we can assume that the first-order deformation $\omega_{1}$ is truncated at the $\omega_{1}^{(2)}$, or in other words we have
\begin{equation}
\begin{aligned}
\omega=\omega_{1}^{(0)} +\omega_{1}^{(1)} +\omega_{1}^{(2)}
\end{aligned}
\end{equation}
thus the (3.14) turns out to be
\begin{equation}
\begin{aligned}
\gamma\omega_{1}^{(2)}=0,\quad \delta\omega_{1}^{(2)}+\gamma\omega_{1}^{(1)}=\partial_{\mu}k^{\mu(1)}_{1},\quad \delta\omega_{1}^{(1)}+\gamma\omega_{1}^{(0)}=\partial_{\mu}k^{\mu(0)}_{1}
\end{aligned}
\end{equation}
from the fact $\gamma\omega_{1}^{(2)}=0$ and since the antighost number of $\omega_{1}^{(2)}$ is $2$, it is reasonable to suppose $\omega_{1}^{(2)}$ in the general form of
\begin{equation}
\begin{aligned}
\omega_{1}^{(2)}=\frac{1}{2}f^{a}_{bc}\eta^{*}_{a}\eta^{b}\eta^{c}
\end{aligned}
\end{equation}
here the constant $f^{a}_{bc}$ is antisymmetric with respect to the indices $b$ and $c$, that is
\begin{equation}
\begin{aligned}
f^{a}_{bc}=-f^{a}_{cb}
\end{aligned}
\end{equation}
 under this assumption, we deduce
\begin{equation}
\begin{aligned}
\delta\omega_{1}^{(2)}=\frac{1}{2}\partial_{\mu}(-f^{a}_{bc}A^{*\mu}_{a}\eta^{b}\eta^{c})+\gamma(f^{a}_{bc}A^{*\mu}_{a}\eta^{b}A^{c}_{\mu})
\end{aligned}
\end{equation}
comparing (3.16) to (3.19), the $\omega_{1}^{(1)}$ is given by
\begin{equation}
\begin{aligned}
\omega_{1}^{(1)}=-f^{a}_{bc}A^{*\mu}_{a}\eta^{b}A^{c}_{\mu}
\end{aligned}
\end{equation}
which further leads to
\begin{equation}
\begin{aligned}
\delta\omega_{1}^{(1)}=&f^{a}_{bc}(-\partial_{\nu}F^{\mu\nu}_{a}-2\partial_{\nu}\partial^{\mu}\partial_{i}F_{a}^{i\nu}+2\partial^{j}\partial_{j}\partial_{i}F_{a}^{i\mu})\eta^{b}A^{c}_{\mu}
\end{aligned}
\end{equation}

On the other hand, it is obvious to check that due to the antisymmetry of the strength $F_{\mu\nu}^{a}$ with respects to $\mu,\nu$, we simply get
\begin{equation}
\begin{aligned}
\gamma F_{\mu\nu}^{a}=0
\end{aligned}
\end{equation}
and keeping this in mind, a straightforward calculation shows
\begin{equation}
\begin{aligned}
&\gamma(\frac{1}{2}f^{a}_{bc}F^{\mu\nu}_{a}A^{b}_{\mu}A^{c}_{\nu}-2f^{a}_{bc}\partial_{i}F^{i\nu}_{a}\partial^{j}(A_{j}^{b}A_{\nu}^{c})-2f^{a}_{bc}\partial_{i}F^{i\nu}_{a}A^{jb}F_{j\nu}^{c})\\
\simeq&f^{a}_{bc}F^{\mu\nu}_{a}\partial_{\mu}\eta^{b}A^{c}_{\nu}+2f^{a}_{bc}\partial^{j}\partial_{i}F^{i\nu}_{a}(\partial_{j}\eta^{b}A_{\nu}^{c}+A_{j}^{b}\partial_{\nu}\eta^{c})-2f^{a}_{bc}\partial_{i}F^{i\nu}_{a}\partial^{j}\eta^{b}F_{j\nu}^{c}\\
\simeq&-f^{a}_{bc}\eta^{b}\partial_{\mu}(F^{\mu\nu}_{a}A^{c}_{\nu})-2f^{a}_{bc}\eta^{b}(\partial_{j}(\partial^{j}\partial_{i}F^{i\nu}_{a}A_{\nu}^{c})-\partial_{\nu}(\partial^{j}\partial_{i}F^{i\nu}_{a}A_{j}^{c})-\partial^{j}(\partial_{i}F^{i\nu}_{a}F_{j\nu}^{c}))\\
=&-f^{a}_{bc}\eta^{b}\partial_{\mu}F^{\mu\nu}_{a}A^{c}_{\nu}-f^{a}_{bc}\eta^{b}F^{\mu\nu}_{a}\partial_{\mu}A^{c}_{\nu}-2f^{a}_{bc}\eta^{b}(\partial_{j}\partial^{j}\partial_{i}F^{i\nu}_{a}A_{\nu}^{c}-\partial_{\nu}\partial^{j}\partial_{i}F^{i\nu}_{a}A_{j}^{c})\\
&-2f^{a}_{bc}\eta^{b}\partial^{j}\partial_{i}F^{i\nu}_{a}F_{j\nu}^{c}+2f^{a}_{bc}\eta^{b}\partial^{j}(\partial_{i}F^{i\nu}_{a}F_{j\nu}^{c})\\
=&f^{a}_{bc}\eta^{b}(\partial_{\nu}F^{\mu\nu}_{a}A^{c}_{\mu}-2\partial_{j}\partial^{j}\partial_{i}F^{i\nu}_{a}A_{\nu}^{c}+2\partial_{\nu}\partial^{j}\partial_{i}F^{i\nu}_{a}A_{j}^{c}-\frac{1}{2}F^{\mu\nu}_{a}F^{c}_{\mu\nu}+2\partial_{i}F^{i\nu}_{a}\partial^{j}F_{j\nu}^{c})
\end{aligned}
\end{equation}
here $\simeq$ denotes the equivalence up to a total derivative term which could be absorbed into the definition of current of $k_{1}^{\mu(0)}$ as we can see from (3.16). Now for the purpose of obtaining consistent deformation terms in $S_{0}\left[A^{a}_{\mu},A^{*\mu}_{a},\eta^{a}\right]$, it is necessary to emphasize that the sum of $\delta\omega_{1}^{(1)}$ and $\gamma\omega_{1}^{(0)}$ should be total derivative terms, in the meantime, we must also demand
\begin{equation}
\begin{aligned}
f^{a}_{bc}\eta^{b}F^{\mu\nu}_{a}F^{c}_{\mu\nu}=0,\quad \quad\quad \quad f^{a}_{bc}\eta^{b}\partial_{i}F^{i\nu}_{a}\partial^{j}F_{j\nu}^{c}=0
\end{aligned}
\end{equation}
since these terms boil down the requirement of the total derivative terms of the sum between (3.21) and (3.23) which result in the extra antisymmetric condition
\begin{equation}
\begin{aligned}
f^{a}_{bc}=-f^{c}_{ba}
\end{aligned}
\end{equation}
for the indices $a,c$. At this point, the solution of (3.16) takes the concrete form of
\begin{equation}
\begin{aligned}
\omega_{1}^{(0)}=\frac{1}{2}f^{a}_{bc}F^{\mu\nu}_{a}A^{b}_{\mu}A^{c}_{\nu}-2f^{a}_{bc}\partial_{i}F^{i\nu}_{a}\partial^{j}(A_{j}^{b}A_{\nu}^{c})-2f^{a}_{bc}\partial_{i}F^{i\nu}_{a}A^{jb}F_{j\nu}^{c}
\end{aligned}
\end{equation}
then putting all of these pieces $\omega_{1}^{(i)}$ together, we have completely obtained the first-order deformation term $S_{1}$ of the deformed master action $S$ as
\begin{equation}
\begin{aligned}
S_{1}=&\int d^{4}x(\frac{1}{2}f^{a}_{bc}F^{\mu\nu}_{a}A^{b}_{\mu}A^{c}_{\nu}-2f^{a}_{bc}\partial_{i}F^{i\nu}_{a}\partial^{j}(A_{j}^{b}A_{\nu}^{c})-2f^{a}_{bc}\partial_{i}F^{i\nu}_{a}A^{jb}F_{j\nu}^{c}\\
&-f^{a}_{bc}A^{*\mu}_{a}\eta^{b}A^{c}_{\mu}+\frac{1}{2}f^{a}_{bc}\eta^{*}_{a}\eta^{b}\eta^{c})
\end{aligned}
\end{equation}

Subsequently, in the consideration of the second-order deformation, let us rewrite the deformation equation $(S_{1},S_{1})+2(S_{0},S_{2})=0$ in the form of local functional
\begin{equation}
\begin{aligned}
s_{11}+2s\omega_{2}=\partial_{\mu}k^{\mu}_{2}
\end{aligned}
\end{equation}
here $(S_{i},S_{j})=\int d^{4}xs_{ij}$ and for convenience we divide $s_{11}$ into three parts $s_{11}=s_{11}^{(0)}+s_{11}^{(1)}+s_{11}^{(2)}$ according to the antighost number, or in detail using the canonical relations
\begin{equation}
\begin{aligned}
(A_{\mu}^{a}(x),A_{b}^{*\nu}(y))&=(A_{b}^{*\nu}(y),A_{\mu}^{a}(x))=-\delta^{a}_{b}\delta^{\nu}_{\mu}\delta^{4}(x-y),\\
(\eta^{a}(x),\eta_{b}^{*}(y))&=(\eta_{b}^{*}(y),\eta^{a}(x))=-\delta^{a}_{b}\delta^{4}(x-y)
\end{aligned}
\end{equation}
and after a careful but direct computation we have
\begin{equation}
\begin{aligned}
s_{11}^{(1)}&=-(f^{a}_{ib}f^{i}_{cd}+f^{a}_{ic}f^{i}_{db}+f^{a}_{id}f^{i}_{bc})A^{*\mu}_{a}\eta^{b}\eta^{c}A^{d}_{\mu},\\
s_{11}^{(2)}&=-\frac{1}{6}(f^{a}_{ib}f^{i}_{cd}+f^{a}_{ic}f^{i}_{db}+f^{a}_{id}f^{i}_{bc})\eta^{*}_{a}\eta^{b}\eta^{c}\eta^{d}
\end{aligned}
\end{equation}
similarly, in order to investigate the solution to (3.28), we develop $\omega_{2}$ as
\begin{equation}
\begin{aligned}
\omega_{2}=\omega_{2}^{(0)} +\omega_{2}^{(1)} +\omega_{2}^{(2)}
\end{aligned}
\end{equation}
here $\mathrm{agh}(\omega_{2}^{(i)})=i$. In this way, the equation (3.28) turns out to be
\begin{equation}
\begin{aligned}
&s_{11}^{(2)}+2\gamma\omega_{2}^{(2)}=\partial_{\mu}k^{\mu(2)}_{2},\\
&s_{11}^{(1)}+2\delta\omega_{2}^{(2)}+2\gamma\omega_{2}^{(1)}=\partial_{\mu}k^{\mu(1)}_{2},\\ &s_{11}^{(0)}+2\delta\omega_{2}^{(1)}+2\gamma\omega_{2}^{(0)}=\partial_{\mu}k^{\mu(0)}_{2}
\end{aligned}
\end{equation}
up to total derivative terms. By inspecting every term on both sides of (3.32), we observe that the sum of $s_{11}^{(2)}$ and $2\gamma\omega_{2}^{(2)}$ can not be a total derivative term, therefore they should vanish, namely
\begin{equation}
\begin{aligned}
s_{11}^{(2)}=0, \quad \quad \omega_{2}^{(2)}=0
\end{aligned}
\end{equation}
and an interesting result will emerge from this
\begin{equation}
\begin{aligned}
&f^{a}_{ib}f^{i}_{cd}+f^{a}_{ic}f^{i}_{db}+f^{a}_{id}f^{i}_{bc}=0
\end{aligned}
\end{equation}
such relation is the usual Jacobi identity which will play a key role in our discussion below, since it allows us to formulate the deformation terms at different orders of the deformed master action in a relatively elegant and compact form. Analogously, substituting the $\omega_{2}^{(2)}=0$ into (3.32) and a similar analysis shows that the only solution for $s_{11}^{(1)},\omega_{2}^{(1)}$ is
\begin{equation}
\begin{aligned}
s_{11}^{(1)}=0, \quad \quad \omega_{2}^{(1)}=0
\end{aligned}
\end{equation}
the deformation equation (3.28) is simplified drastically with the help of the identity (3.33) and (3.35) that the $s_{11}^{(1)},s_{11}^{(2)},\omega_{2}^{(1)},\omega_{2}^{(2)}$ do not show up in (3.28). Therefore, we just need to solve the third equation in (3.32) which means only $s_{11}^{(0)},\omega_{2}^{(0)}$ terms survive in $s_{11}$ and $\omega_{2}$ respectively.

Remembering that the Jacobi identity (3.34) and relations (3.18), for subsequent discussions, it is convenient for us to recast the above expressions into the simple matrix form. To do so, let us introduce a basis of matrix generators $(\Gamma_{a})_{ij}=\Gamma_{aj}^{i}$ satisfying
\begin{equation}
\begin{aligned}
\left[\Gamma_{b},\Gamma_{c}\right]=f^{a}_{bc}\Gamma_{a}
\end{aligned}
\end{equation}
and we also require the coefficients $\Gamma_{aj}^{i}$ normalized as $\mathrm{tr}(\Gamma_{a}\Gamma_{b})=\delta_{ab}$, of course these automatically fulfill the Jacobi identity
\begin{equation}
\begin{aligned}
\left[\Gamma_{a},\left[\Gamma_{b},\Gamma_{c}\right]\right]+\left[\Gamma_{b},\left[\Gamma_{c},\Gamma_{a}\right]\right]+\left[\Gamma_{c},\left[\Gamma_{a},\Gamma_{b}\right]\right]=0
\end{aligned}
\end{equation}
and noting that in the mathematical literature, these $\Gamma_{a}$ constitute the generators of some Lie algebra $\mathfrak{g}$ and thus from now on, we can take the gauge fields $A_{\mu}^{a}$, strength fields $F^{a}_{\mu\nu}$ and the ghost fields/antifields $\eta^{a},\eta^{*}_{a}$ in the Lie algebra $\mathfrak{g}$-valued form, that is we do the replacement
\begin{equation}
\begin{aligned}
A^{a}_{\mu}\rightarrow A_{\mu}=A^{a}_{\mu}\Gamma_{a},\quad  F^{a}_{\mu\nu}\rightarrow F_{\mu\nu}=F^{a}_{\mu\nu}\Gamma_{a},\quad \eta^{a}\rightarrow \eta=\eta^{a}\Gamma_{a},\quad \eta^{*}_{a}\rightarrow \eta^{*}=\eta^{*}_{a}\Gamma_{a}
\end{aligned}
\end{equation}
as a result, in terms of the matrix generators $\Gamma_{a}$, we are capable of rewriting the first-order deformation $S_{1}$ in a more concise way
\begin{equation}
\begin{aligned}
S_{1}=\mathrm{tr}\int d^{4}x&(\frac{1}{2}F^{\mu\nu}\left[A_{\mu},A_{\nu}\right]-2\partial_{i}F^{i\nu}\partial^{j}\left[A_{j},A_{\nu}\right]-2\partial_{i}F^{i\nu}\left[A^{j},F_{j\nu}\right]\\
&-A^{*\mu}\left[\eta,A_{\mu}\right]+\frac{1}{2}\eta^{*}\left[\eta,\eta\right])
 \end{aligned}
\end{equation}
remarkablely, we stress here that this compact form of the deformation terms will make the expressions of our results less cumbersome and more intuitive.

\subsection{Higher-order deformation}
We turn back into the deformation equations (3.10) and our aim in this section is to find appropriate higher-order deformation terms which satisfy these deformation equations. First of all, for the sake of brevity we divide $S_{1}$ into two sectors
\begin{equation}
\begin{aligned}
S_{1}^{g}&=\mathrm{tr}\int d^{4}x(\frac{1}{2}F^{\mu\nu}\left[A_{\mu},A_{\nu}\right]-2\partial_{i}F^{i\nu}\partial^{j}\left[A_{j},A_{\nu}\right]-2\partial_{i}F^{i\nu}\left[A^{j},F_{j\nu}\right]),\\
S_{1}^{A}&=\mathrm{tr}\int d^{4}x(-A^{*\mu}\left[\eta,A_{\mu}\right]+\frac{1}{2}\eta^{*}\left[\eta,\eta\right])
 \end{aligned}
\end{equation}
here $S_{1}^{g}$ is the pure gauge part and $S_{1}^{A}$ involves the antifields terms only. Under such decomposition and using Jacobi identity, it is easy to establish the following deformation equation
\begin{equation}
\begin{aligned}
(S_{1}^{A},S_{1}^{g})+sS_{2}=0
\end{aligned}
\end{equation}
 the solution of above equality takes the form of
\begin{equation}
\begin{aligned}
S_{2}=&\mathrm{tr}\int d^{4}x(-\frac{1}{4}\left[A_{\mu},A_{\nu}\right]\left[A^{\mu},A^{\nu}\right]+\partial _{\mu}\left[A^{\mu},A^{\nu}\right]\partial^{i}\left[A_{i},A_{\nu}\right]+2\partial _{\mu}\left[A^{\mu},A^{\nu}\right]\left[A^{i},F_{i\nu}\right]\\
&+\left[A_{\mu},F^{\mu\nu}\right]\left[A^{i},F_{i\nu}\right]+2\partial _{\mu}F^{\mu\nu}\left[A^{i},\left[A_{i},A_{\nu}\right]\right])
\end{aligned}
\end{equation}
this solution can be examined by a length but direct calculations with the aid of Jacobi identity. Similarly, as far as the third-order deformation term is concerned, we set up the following deformation equation
\begin{equation}
\begin{aligned}
(S_{1}^{A},S_{2})+sS_{3}=0
\end{aligned}
\end{equation}
which can be worked out as
\begin{equation}
\begin{aligned}
S_{3}=&\mathrm{tr}\int d^{4}x(-2\partial _{\mu}\left[A^{\mu},A^{\nu}\right]\left[A^{i},\left[A_{i},A_{\nu}\right]\right]-2\left[A_{\mu},F^{\mu\nu}\right]\left[A^{i},\left[A_{i},A_{\nu}\right]\right])\\
\end{aligned}
\end{equation}
to proceed further, we focus the attentions on the fourth-order deformation $S_{4}$ at order $g^{4}$ and because of the fact $(S_{2},S_{2})=0$, the deformation equation $2(S_{0},S_{4})+2(S_{1},S_{3})+(S_{2},S_{2})=0$ turns out to be
\begin{equation}
\begin{aligned}
(S_{1}^{A},S_{3})+sS_{4}=0
\end{aligned}
\end{equation}
in this manner, we have the solution
\begin{equation}
\begin{aligned}
S_{4}=&\mathrm{tr}\int d^{4}x(\left[A_{\mu},\left[A^{\mu},A^{\nu}\right]\right]\left[A^{i},\left[A_{i},A_{\nu}\right]\right])
\end{aligned}
\end{equation}

Now we are in the position to determine the fifth-order deformation in the deformed master equations (3.10), again taking advantage of the Jacobi identity, it is not hard to check that
\begin{equation}
\begin{aligned}
(S_{1},S_{4})=0
\end{aligned}
\end{equation}
while $(S_{2},S_{3})=0$ is fairly obvious, thus it follows from (3.10) that the consistency at order $g^{5}$ implies $S_{5}=0$ and subsequently all other deformations with orders higher than four are equal to zero, $S_{i}=0$ for $i\geq5$ precisely. To conclude, the solution of the deformed master equation, consistent to all orders of the deformation parameter $g$ is given by
\begin{equation}
\begin{aligned}
S=S_{0}+S_{1}+S_{2}+S_{3}+S_{4}
\end{aligned}
\end{equation}
to simplify this expression, let us introduce the field strength $\mathcal{F}_{\mu\nu}$ together with the covariant derivative $D_{\mu}$
\begin{equation}
\begin{aligned}
\mathcal{F}_{\mu\nu}=F_{\mu\nu}-g\left[A_{\mu},A_{\nu}\right],\quad \quad D_{\mu}=\partial_{\mu}-g\left[A_{\mu},\quad \right]
\end{aligned}
\end{equation}
the antighost number zero part $\tilde{S}_{0}$ of the deformed master action $S$ should provide the Lagrangian action of the interacting theory which can be expressible in the manner of
\begin{equation}
\begin{aligned}
\tilde{S}_{0}\left[A_{\mu}\right]=\mathrm{tr}\int d^{4}x(-\frac{1}{4}\mathcal{F}_{\mu\nu}\mathcal{F}^{\mu\nu}+D_{\mu}\mathcal{F}^{\mu\nu}D^{\lambda}\mathcal{F}_{\lambda\nu})
\end{aligned}
\end{equation}
it is essential to remark here that the deformed solution (3.48) contains all the information on the gauge structure of the resulting interacting theory. Moreover, evidently the system (3.50) is invariant under the corresponding deformed transformations
\begin{equation}
\begin{aligned}
\delta_{\varepsilon}A_{\mu}^{a}=\partial_{\mu}\varepsilon^{a}-gf^{a}_{bc}A_{\mu}^{b}\varepsilon^{c}
\end{aligned}
\end{equation}
for arbitrary functions $\varepsilon^{a}$. By comparing the original transformation (3.3), we find that the deformation procedure modifies the gauge transformation and in addition, the (3.51) is the usual local non-Abelian gauge transformation if we interpret the deformation parameter $g$ as the coupling constant among gauge fields.

The above consequence enables us to formulate the total solution of the deformed master action including the non-minimal term in the form of
\begin{equation}
\begin{aligned}
S=&\tilde{S}_{0}+\mathrm{tr}\int d^{4}x(A^{*\mu}\partial_{\mu}\eta-gA^{*\mu}\left[\eta,A_{\mu}\right]+\frac{1}{2}g\eta^{*}\left[\eta,\eta\right]+\bar{\eta}^{*}\lambda)
\end{aligned}
\end{equation}
for the purpose of calculating the path integral of this non-Abelian coupling system at the quantum level, analogously, an appropriate gauge-fixing fermion has to be selected as
\begin{equation}
\begin{aligned}
\Psi=\mathrm{tr}\int d^{3}x\bar{\eta}(\partial^{\mu}A_{\mu}+\frac{\square}{m^{2}}\partial^{\mu}A_{\mu}-\frac{\alpha}{2}\lambda)
\end{aligned}
\end{equation}
in this way, one obtains the following identifications of the non-trivial values of the antighost fields
\begin{equation}
\begin{aligned}
\bar{\eta}^{*}=\frac{\delta\Psi}{\delta\bar{\eta}}=-(\partial^{\mu}A_{\mu}+\frac{\square}{m^{2}}\partial^{\mu}A_{\mu}-\frac{\alpha}{2}\lambda),\quad \quad \quad A^{\ast\mu}=\frac{\delta\Psi}{\delta A_{\mu}}=\partial^{\mu}\bar{\eta}+\frac{\square}{m^{2}}\partial^{\mu}\bar{\eta}
\end{aligned}
\end{equation}
plugging these relations back into (3.52) and after integrating out the auxiliary field $\lambda$, we simply get the transition-amplitude of the non-Abelian system (3.50)
 \begin{equation}
\begin{aligned}
Z=\int\left[DA_{\mu}\right]\left[D\eta\right]\left[D\bar{\eta}\right]\mathrm{exp}\frac{i}{\hbar}(\tilde{S}_{0}+\mathrm{tr}\int d^{4}x(-\frac{1}{2\alpha}\left[(1+\frac{\square}{m^{2}})\partial^{\mu}A_{\mu}\right]^{2}-\partial^{\mu}\bar{\eta}(1+\frac{\square}{m^{2}})\bar{D}_{\mu}\eta))
\end{aligned}
\end{equation}
here $\bar{D}_{\mu}=\partial_{\mu}+g\left[A_{\mu},\quad \right]$ and this formula precisely coincides with the expression in ~\cite{14} derived from the constructions of BRST charge and BRST-invariant Hamiltonian in BFV quantization method. Obviously, as we can see, the path integral in BV formalism shows to be more convenient and efficient than within the Hamiltonian BRST formalism, though these two methods are equivalent.

\section{Resulting interacting theory}
\subsection{Degrees of freedom}
Let us turn the attentions to the calculations of the number of physical degrees of freedom of the resulting interacting theory (3.50) and we proceed using the standard Ostrogradski formalism. To implement this method, for the components of every index $a$, we define the corresponding canonical momenta for the independent dynamical variables $(A_{\mu}^{a},\Gamma_{\mu}^{a}=\dot{A}_{\mu}^{a})$
\begin{equation}
\begin{aligned}
\pi^{\mu a}=\frac{\partial\mathcal{L}}{\partial \dot{A}_{\mu}^{a}}-\partial_{\nu}\frac{\partial\mathcal{L}}{\partial(\partial_{\nu} \dot{A}_{\mu}^{a})},\quad \quad \quad \phi^{\mu a}=\frac{\partial\mathcal{L}}{\partial \dot{\Gamma}_{\mu}^{a}}
\end{aligned}
\end{equation}
after a straightforward computation, we simply get
\begin{equation}
\begin{aligned}
\pi^{ia}&=\mathcal{F}^{i0}+\frac{1}{m^{2}}D_{i}D_{\lambda}\mathcal{F}^{\lambda0}-\frac{1}{m^{2}}D_{0}D_{\lambda}\mathcal{F}^{\lambda i},\quad \quad\quad \pi^{0a}=\frac{1}{m^{2}}D_{i}D_{\lambda}\mathcal{F}^{\lambda i},\\
\phi^{ia}&=\frac{1}{m^{2}}D_{\lambda}\mathcal{F}^{\lambda i},\quad \quad \quad \quad \quad \quad  \phi^{0a}=0
\end{aligned}
\end{equation}
together with the standard Poisson brackets
\begin{equation}
\begin{aligned}
\{A_{\mu}^{a}(x),\pi^{\nu b}(y)\}=\delta^{ab}\delta_{\mu}^{\nu}\delta^{3}(x-y),\quad \quad \{\Gamma_{\mu}^{a}(x),\phi^{\nu b}(y)\}=\delta^{ab}\delta_{\mu}^{\nu}\delta^{3}(x-y)
\end{aligned}
\end{equation}
a usual Legendre transform immediately yields the canonical Hamiltonian in the form of
\begin{equation}
\begin{aligned}
H_{c}=&\mathrm{tr}\int d^{3}x(\pi^{\mu}\dot{A}_{\mu}+\phi^{\mu}\dot{\Gamma}_{\mu}-\mathcal{L})\\
=&\mathrm{tr}\int d^{3}x(\pi^{0}\Gamma_{0}+\pi^{i}\Gamma_{i}+\phi^{i}(\frac{m^{2}}{2}\phi_{i}+D_{i}\Gamma_{0}+D^{j}\mathcal{F}_{ij}+2g\left[A_{0},\Gamma_{i}\right]-g\left[A_{0},\partial_{i}A_{0}+g\left[A_{0},A_{i}\right]\right])\\
&+\frac{1}{4}\mathcal{F}^{ij}\mathcal{F}_{ij}-\frac{1}{2}(\Gamma_{i}-\partial_{i}A_{0}-g\left[A_{0},A_{i}\right])^{2}-\frac{1}{2m^{2}}(D^{i}(\Gamma_{i}-\partial_{i}A_{0}-g\left[A_{0},A_{i}\right]))^{2})
\end{aligned}
\end{equation}

It is easy to determine the primary constraints $\Phi_{1}^{a}\equiv \phi^{0a}\approx 0$ in (4.2) and the primary Hamiltonian is given by
\begin{equation}
\begin{aligned}
H_{T}=H_{c}+\mathrm{tr}\int d^{3}xu_{1}\Phi_{1}
\end{aligned}
\end{equation}
here the $u_{1}$ is the Lagrange multipliers and when varied independently, it ensures the primary constraints. Furthermore, the preservation of the primary constraint over time is acquired through the Poisson brackets between $\Phi_{1}^{a}$ with the primary Hamiltonian which leads to the secondary constraints
\begin{equation}
\begin{aligned}
\Phi_{2}^{a}\equiv (\pi_{0}-D_{i}\phi^{i})^{a}\approx 0
\end{aligned}
\end{equation}
while conserving $\Phi_{2}^{a}$ in time
\begin{equation}
\begin{aligned}
0\approx\dot{\Phi}_{3}^{a}=\{\Phi_{2}^{a},H_{T}\}
\end{aligned}
\end{equation}
a collection of new constraints will emerge and in order to acquire their exact expressions, by means of the canonical commutative relations we have
\begin{equation}
\begin{aligned}
&\{\pi^{a}_{0},H_{T}\}=(g(\left[\partial_{i}A_{0},\phi^{i}\right]-\partial_{i}\left[\phi^{i},A_{0}\right])-2g\left[\Gamma_{i},\phi^{i}\right]+g^{2}(\left[\left[A_{0},A_{i}\right],\phi^{i}\right]+\left[A_{i},\left[\phi^{i},A_{0}\right]\right]))^{a}
\end{aligned}
\end{equation}
as well as
\begin{equation}
\begin{aligned}
&\{(-D_{i}\phi^{i})^{a},H_{T}\}=(D_{i}\pi^{i}+g\left[\Gamma_{i},\phi^{i}\right]+2gD_{i}\left[\phi^{i},A_{0}\right])^{a}
\end{aligned}
\end{equation}
 combining (4.8) and (4.9) together, it is immediately to get the tertiary constraints
 \begin{equation}
\begin{aligned}
&\Phi^{a}_{3}\equiv\{(\pi_{0}-D_{i}\phi^{i}),H_{T}\}=(D_{i}\pi^{i}-g\left[\Gamma_{i},\phi^{i}\right]-g\left[A_{0},D_{i}\phi^{i}\right])^{a}
 \end{aligned}
\end{equation}
afterwards, we will proceed to the determination of the new constraints from the conservation equations
\begin{equation}
\begin{aligned}
0\approx\dot{\Phi}_{3}^{a}=\{\Phi_{3}^{a},H_{T}\}
\end{aligned}
\end{equation}
and there is no difficulty in carrying out
\begin{equation}
\begin{aligned}
\{(D_{i}\pi^{i})^{a},H_{T}\}&=(-g\left[\Gamma_{i},\pi^{i}\right]+gD^{j}\left[\mathcal{F}_{ij},\phi^{i}\right]+g^{2}D_{i}\left[\left[\phi^{i},A_{0}\right],A_{0}\right])\\
\end{aligned}
\end{equation}
a similar but direct calculation gives us
\begin{equation}
\begin{aligned}
\{(-g\left[A_{0},D_{i}\phi^{i}\right])^{a},H_{T}\}&=(g\left[A_{0},D_{i}\pi^{i}\right]+g^{2}\left[A_{0},\left[\Gamma_{i},\phi^{i}\right]\right]+2g^{2}\left[A_{0},D_{i}\left[\phi^{i},A^{0}\right]\right])^{a}
\end{aligned}
\end{equation}
together with
\begin{equation}
\begin{aligned}
\{(-g\left[\Gamma_{i},\phi^{i}\right])^{a},H_{T}\}&=(g\left[\Gamma_{i},\pi^{i}\right]-gD^{j}\left[\mathcal{F}_{ij},\phi^{i}\right]+g^{2}\left[\left[A_{0},D_{i}A_{0}\right],\phi^{i}\right]-2g^{2}\left[A_{0},\left[\Gamma_{i},\phi^{i}\right]\right])^{a}\\
\end{aligned}
\end{equation}
then a careful summation works out
\begin{equation}
\begin{aligned}
&D_{i}(\left[\left[\phi^{i},A_{0}\right],A_{0}\right])+2\left[A_{0},D_{i}\left[\phi^{i},A^{0}\right]\right]+\left[\left[A_{0},D_{i}A_{0}\right],\phi^{i}\right]\\
=&\left[\left[\phi^{i},A_{0}\right],D_{i}A_{0}\right]+\left[A_{0},D_{i}\left[\phi^{i},A^{0}\right]\right]+\left[\left[A_{0},D_{i}A_{0}\right],\phi^{i}\right]\\
=&\left[\left[\phi^{i},A_{0}\right],D_{i}A_{0}\right]+\left[A_{0},\left[D_{i}\phi^{i},A^{0}\right]\right]+\left[A_{0},\left[\phi^{i},D_{i}A^{0}\right]\right]+\left[\left[A_{0},D_{i}A_{0}\right],\phi^{i}\right]\\
=&-\left[A_{0},\left[A^{0},D_{i}\phi^{i}\right]\right]
\end{aligned}
\end{equation}
putting all of these results together, we succeeded in gaining the new constraints as follows
\begin{equation}
\begin{aligned}
\Phi^{a}_{4}\equiv\{\Phi^{a}_{3},H_{T}\}=g(\left[A_{0},D_{i}\pi^{i}-g\left[\Gamma_{i},\phi^{i}\right]-g\left[A_{0},D_{i}\phi^{i}\right]\right])^{a}=g(\left[A_{0},\Phi_{3}\right])^{a}\approx 0\\
\end{aligned}
\end{equation}
obviously, no additional constraints are obtained from the consistency of the constraints $\Phi^{a}_{3}$ and the iterative process stops here.

In the current context, making using of the Jacobi identity, we observe that every $\Phi_{i}^{a}$ has vanishing brackets with all the other constraints and hence there are three first-class constraints for every index $a$. In conclusion, the number of physical degrees of freedom of the resulting interacting theory is equal to
\begin{equation}
\begin{aligned}
\mathcal{N}=(16-2\times3)N/2=5N
\end{aligned}
\end{equation}
which coincides with the number of physical degrees of freedom of the free higher derivative system (3.1). This can happen, since the inclusion of consistent interactions should not change the number of degrees of freedom which also can be seen by comparing the number of first-class constraints in two different systems.

\subsection{Stability}

Finally, let us provide a simple explanation on the issues of stability in the non-Abelian higher derivative system and for convenience, it is better to divide the resulting Lagrangian into two parts
\begin{equation}
\begin{aligned}
\tilde{S}_{0}=\mathrm{tr}\int d^{4}x&(-\frac{1}{4}F_{\mu\nu}F^{\mu\nu}+\frac{1}{2m^{2}}\partial_{\mu}F^{\mu\nu}\partial^{\lambda}F_{\lambda\nu}+U(A))
\end{aligned}
\end{equation}
here the $U(A)$ stands for the self-interacting terms among the gauge fields and the equation of motion of the gauge fields $A_{\mu}$ takes the form of
\begin{equation}
\begin{aligned}
(W(W+m^{2}))_{\mu\nu}A_{\nu}+m^{2}U^{'}_{\mu}(A)=0
\end{aligned}
\end{equation}
here $U^{'}_{\mu}$ denotes the Euler-Lagrange derivative of $U$ with respect to $A_{\mu}$
\begin{equation}
\begin{aligned}
U^{'}_{\mu}(A)=\sum_{k=0}^{2}(-1)^{k}\frac{\partial^{k}}{\partial x^{\mu_{1}}...\partial x^{\mu_{k}}}\frac{\partial U}{\partial(\partial_{\mu_{1}}...\partial_{\mu_{k}}A_{\mu})}
\end{aligned}
\end{equation}
then in a similar way, it is necessary to introduce a suitable set of dynamic fields
\begin{equation}
\begin{aligned}
\eta_{1}=-\frac{1}{m^{2}}WA,\quad \quad \eta_{2}=\frac{1}{m^{2}}(W+m^{2})A
\end{aligned}
\end{equation}
which fulfill the relation
\begin{equation}
\begin{aligned}
\eta_{1}+\eta_{2}=A
\end{aligned}
\end{equation}
here the $\eta_{i}$ are all four-component fields valued in the Lie algebra form. At this stage, let us establish the following action functional which is parameterized by $\alpha,\beta$
\begin{equation}
\begin{aligned}
&S_{1}\left[\alpha,\beta,\eta_{1},\eta_{2}\right]=\mathrm{tr}\int d^{4}x\left[\frac{\alpha}{2}\eta_{1}(W+m^{2})\eta_{1}+\frac{\beta}{2}\eta_{2}W\eta_{2}-U(\alpha\eta_{1}-\beta\eta_{2})\right]
\end{aligned}
\end{equation}
and the equations of motions are given by
\begin{equation}
\begin{aligned}
\frac{\delta S_{1}}{\delta \eta_{1}}&=\alpha(W+m^{2})\eta_{1}-\alpha U^{'}(\alpha\eta_{1}-\beta\eta_{2})=0,\\
\frac{\delta S_{1}}{\delta \eta_{2}}&=\beta W\eta_{2}+\beta U^{'}(\alpha\eta_{1}-\beta\eta_{2})=0
\end{aligned}
\end{equation}
especially, notice that when $\alpha=-\beta=1$, using (4.21), the dynamic equations (4.24) turn out to be (4.19) which implies that in such special case, the two systems are equivalent. Next, as we have already demonstrated previously, once the action (4.23) is invariant under the spacetime translations, the Noether's theorem will produce two-parametric conserved second-rank tensors ~\cite{17,18,19}
\begin{equation}
\begin{aligned}
\Theta^{\mu}_{\nu}=\alpha\Theta^{\mu}_{\nu}(\eta_{1})+\beta\Theta^{\mu}_{\nu}(\eta_{2})+(\Theta^{\mu}_{\nu})_{int}(\alpha\eta_{1}-\beta\eta_{2})
\end{aligned}
\end{equation}
here the term $(\Theta^{\mu}_{\nu})_{int}$ denotes the energy-momentum tensors associated with the self-interactions and the quantities $\Theta^{\mu}_{\nu}(\eta_{i})$ are the energy-momentum tensors of the free theories satisfying
\begin{equation}
\begin{aligned}
(W+m^{2})\eta_{1}=0,\quad \quad \quad W\eta_{2}=0
\end{aligned}
\end{equation}
or more explicitly
\begin{equation}
\begin{aligned}
\Theta^{\mu}_{\nu}(\eta_{1})&=\mathrm{tr}(\frac{1}{4}\delta^{\mu}_{\nu}F^{1}_{\rho\lambda}F^{\rho\lambda}_{1}-F^{1}_{\nu\lambda}F^{\mu\lambda}_{1}-\eta_{1\nu}\partial_{\omega}F^{\omega\mu}_{1}-\frac{1}{2}m^{2}\delta^{\mu}_{\nu}\eta_{1\omega}\eta^{\omega}_{1}),\\
\Theta^{\mu}_{\nu}(\eta_{2})&=\mathrm{tr}(\frac{1}{4}\delta^{\mu}_{\nu}F_{\rho\lambda}^{2}F^{\rho\lambda}_{2}-F^{2}_{\nu\lambda}F^{\mu\lambda}_{2}-\eta_{2\nu}\partial_{\omega}F^{\omega\mu}_{2})
\end{aligned}
\end{equation}
again here we define
\begin{equation}
\begin{aligned}
F^{\rho\lambda}_{i}=\partial^{\rho}\eta^{\lambda}_{i}-\partial^{\lambda}\eta^{\rho}_{i},\quad \quad \quad i=1,2
\end{aligned}
\end{equation}
for the case of our interest, taking advantage of (4.26), the 00-component of the above conserved energy-momentum tensors becomes
\begin{equation}
\begin{aligned}
\Theta^{0}_{0}=\mathrm{tr}\left[\alpha(\frac{1}{4}F_{\rho\lambda}^{1}F_{\rho\lambda}^{1}+\frac{1}{2}m^{2}\eta_{1\omega}\eta_{1\omega})+\frac{1}{4}\beta F_{\rho\lambda}^{2}F_{\rho\lambda}^{2}\right]+(\Theta^{0}_{0})_{int}(\alpha\eta_{1}-\beta\eta_{2})
\end{aligned}
\end{equation}
analogously, when $\alpha=-\beta=1$, upon substitution of (4.21) into (4.29), we find that this two-parametric conserved quantity includes the energy density of the original higher derivative interacting system (4.18). With these results in hand, if we choose
\begin{equation}
\begin{aligned}
\beta_{1}>0,\quad \quad\quad \beta_{2}>0
\end{aligned}
\end{equation}
both of the free factors for $\eta_{1}$ and $\eta_{2}$ are stable. Now as explained in ~\cite{17,53,54,55}, even if the interaction term $(\Theta^{0}_{0})_{int}$ is not positive definite, the energy can still have a local minimum in a neighborhood of zero solution. These theories with such  "locally stable" behavior ~\cite{53,54,55} are also considered as physically acceptable models which could be studied by means of perturbation expansion.

\section{Conclusion and discussion}

In this paper, we investigate the stability of Podolsky's generalized electrodynamics with the aid of a series of conserved quantities including the standard canonical energy-momentum tensors in the higher derivative theory. These conserved quantities are derived from the higher order symmetries connected with the spacetime translation invariance of the action functional. Especially, the 00-component can be bounded depending on appropriate values of the parameters in the conserved tensors and in this manner, it will stabilize the dynamics of the higher derivative system, though the canonical energy is usually unbounded from below. Then we consider the inclusions of consistent interactions in the Podolsky's model through the deformation viewpoint of the master equations within the framework of BRST antifields approach. By means of the local BRST cohomology and the standard homological perturbative expansion procedure, there is no difficulty for us in obtaining the deformation terms of the extended master action order by order via solving a collection of recursive functional equations. We show that the fifth-order and the corresponding higher-order deformations vanish precisely because of the obstruction of consistent local self-couplings. As we can anticipate, the final expression after the deformation process is the non-Abelian extension of the original generalized electrodynamics formulated in the form of the non-Abelian curvature terms with higher order action of the ordinary derivative replaced by the covariant derivative. In this way, the resulting interacting theory can be regarded as taking values in the fundamental representation of some specific Lie algebra determined from the coefficients $f^{a}_{bc}$ we choose. Also we demonstrate that once the free derived theory is stable, it is also true for the inclusions of interactions of derived models.

There are many interesting and further generalizations of this work. In particular, it is natural to consider this deformation in the case of the gauge fields together with higher derivative matter fields in the free Lagrangian including the massless real scalar fields, the massive complex scalar fields and Dirac spinor fields. The main difference in these situations are the existences of two different types of couplings in the first-order deformation. One is only for the gauge fields and the others involving the matter fields should possess some global invariances which lead to the conserved currents due to the Noether's theorem. Of course, from the consistency of the first-order deformation, we will acquire the minimal couplings between gauge and matter fields after the deformation procedure. The issues of stability both in the free and interacting systems can be analyzed in a similar way along the idea in our discussions. Another application of this paper is to the Chern-Simons theory with higher-order derivative terms with respect to the $U(1)$ gauge fields or the higher-order Maxwell-Chern-Simons-Podolsky theory and the relevant route with the corresponding computations are analogous to our derivations presently. In such system, the primary wave operator is no longer the Maxwell operator of order two but the Chern-Simons operator of order one. In this form, the characteristic polynomial will give rise to four additional conserved tensors and the linear combination of these conserved quantities is in fact a quadratic form in terms of the gauge field with respect to its derivatives.  Now the higher derivative dynamics is stable if these parameters and the coefficients in Lagrangian ensure the positive definite of this quadratic form.

Besides the idea of deformations of the master action in BV formalism, one further direction of this line which will be significant is that we may apply the similar Hamiltonian BRST-invariant method to the Podolsky's higher derivative model. In such situation, taking advantage of the auxiliary fields, we are able to reduce the higher derivative to first order and the number of dynamical variables of the resulting system will be twice than the original one. Afterwards, it is direct to obtain the BRST charge and the BRST-invariant Hamiltonian from the standard method in the procedure of BRST quantization and we should express the deformed Hamiltonian and the BRST charge in terms of power series expansion of the deformation parameter. Then for the purpose of gaining the consistent interactions, it is reasonable to require that the nilpotency of the BRST charge and the commutativity between the Hamiltonian and BRST charge should be preserved which will give rise to a set of iterative equations coming from the perturbative expansion order by order. Through solving these equations recursively, we are able to receive various consistent interaction terms at different orders in the free system for the gauge fields. We might obtain the non-Abelian Lagrangian action by extracting the first-class Hamiltonian of the interacting theory after the deformation as we can imagine. In addition, such Hamiltonian BRST-invariant deformations also can be employed into Podolsky' theory within the Ostrogradsky formalism in a parallel way. All of these would be interesting to exploit in future.

\acknowledgments
The author would like to thank the G.W.Wan for long time encouragements and is grateful to S.M.Zhu for useful support.



\begin{thebibliography}{99}

 \bibitem{1}
A. Anisimov, E. Babichev and A. Vikman,  \emph{B-inflation}, JCAP 0506: 006, (2005).
\bibitem{2}
R.P. Woodard,  \emph{Avoiding Dark Energy with $1/R$ Modifications of Gravity}, Lect. Notes. Phys, 720, 403 (2007).
\bibitem{3}
P. Mukherjee and B.Paul, \emph{Gauge invariances of higher derivative Maxwell-Chern-Simons field theory - a new Hamiltonian approach}, Phys. Rev. D 85,045028 (2011).
\bibitem{4}
M. Crisostomi, R. Klein and D. Roest,\emph{Higher Derivative Field Theories: Degeneracy Conditions and Classes}, JHEP(6), 1-29 (2017).

\bibitem{5}
B. Podolsky,  \emph{A generalized electrodynamics. I. Nonquantum}, Phys. Rev. (2) 62 (1942), 68–71.
\bibitem{6}
B. Podolsky and P. Schwed,\emph{Review of a Generalized Electrodynamics}, Rev. Mod. Phys. 20, 40 (1948).
\bibitem{7}
A.E.S. Green, \emph{Self-energy and interaction energy in Podolsky's generalized electrodynamics}, Phys. Rev. (2) 72 (1947), 628–631.

\bibitem{8}
C.A.P. Galvao and B.M.Pimentel Escobar,\emph{The canonical structure of Podolsky generalized electrodynamics}, Can.J.Phys. 66, 460 (1988).
\bibitem{9}
M.C. Bertin, B.M. Pimentel and G.E.R. Zambrano,\emph{The canonical structure of Podolsky's generalized electrodynamics on the Null-Plane}, J.Math.Phys. 52(10)(2009).
\bibitem{10}
J. Barcelos-Neto, C.A.P. Galvao and C.P. Natividade,\emph{Quantization of Podolsky theory in the BFV formalism},  Z. Phys. C 52, 559 (1991).
\bibitem{11}
R. Bufalo, B.M. Pimentel and G.E.R. Zambrano, \emph{Path Integral Quantization of Generalized Quantum Electrodynamics},	Phys.Rev.D 83, 045007 (2011).
\bibitem{12}
R. Bufalo and B.M. Pimentel,\emph{Batalin-Fradkin-Vilkovisky quantization of the generalized scalar electrodynamics}, Phys.Rev.D 88, 065013 (2013).
\bibitem{13}
A.A. Nogueira, C. Palechor and A.F. Ferrari, \emph{Reduction of order and Fadeev-Jackiw formalism in generalized electrodynamics}, Nucl.Phys. B 939 (2019) 372-390.
\bibitem{14}
R. Bufalo and B. M. Pimentel, \emph{Higher-derivative non-Abelian gauge fields via the Faddeev-Jackiw formalism}, Eur. Phys. J. C 74, 2993 (2014).


\bibitem{15}
M. Ostrogradsky, Mem. Ac. St. Petersbourg VI 4 (1850) 385.
\bibitem{16}
F.J. Urries and J. Julve, \emph{Ostrogradski formalism for higher-derivative scalar field theories}, J.Phys.A 31:6949-6964,1998.


\bibitem{17}
D.S. Kaparulin, S.L. Lyakhovich and A.A. Sharapov,\emph{Classical and quantum stability of higher-derivative dynamics}, Eur. Phys. J. C 74(10),2014.
\bibitem{18}
D.S. Kaparulin,\emph{Conservation Laws and Stability of Field Theories of Derived Type}, Symmetry 2019, 11(5), 642.
\bibitem{19}
V.A. Abakumova, D.S. Kaparulin and S.L. Lyakhovich,\emph{Stable interactions in higher derivative field theories of derived type}, Phys. Rev. D 99, 045020,(2019).
\bibitem{20}
V.A. Abakumova, D.S. Kaparulin and S.L. Lyakhovich,\emph{Conservation laws and stability of higher derivative extended Chern-Simons}, arXiv:1907.02267.	
\bibitem{21}
V.A. Abakumova, D.S. Kaparulin and S.L. Lyakhovich,\emph{Stable interactions between higher derivative extended Chern-Simons and charged scalar field}, 	arXiv:1907.08075.
\bibitem{22}
V.A. Abakumova, D.S. Kaparulin and S.L. Lyakhovich, \emph{Stable Interactions between extended Chern-Simons theory and charged scalar field with higher derivatives: Hamiltonian formalism}, Russ. Phys. J. 62 (2019).


\bibitem{23}
C. Becchi, A. Rouet and R. Stora, \emph{Renormalization of the abelian Higgs-Kibble model}, Commun. Math. Phys. 42:127–162, 1975.
\bibitem{24}
C. Becchi, A. Rouet and R. Stora, \emph{Renormalization of gauge theories}, Ann. Phys. 98(2):287–321, 1976.
\bibitem{25}
I.V. Tyutin, \emph{Gauge Invariance in Field Theory and Statistical Physics in Operator Formalism}, arXiv:0812.0580.
\bibitem{26}
M. Henneaux, \emph{Lectures on the antifield-BRST formalism for gauge theories}, Nucl. Phys.B 18 A (1990) 47-106.
\bibitem{27}
 G. Barnich, F. Brandt and M. Henneaux, \emph{Local BRST cohomology in gauge theories}, Phys. Rept. 338 (2000) 439.
\bibitem{28}
A. Fuster, M. Henneaux and A. Maas, \emph{BRST-antifield Quantization: a Short Review}, Int.J.Geom.Meth.Mod.Phys. 2 (2005) 939-964.
 \bibitem{29}
M. Henneaux and C. Teitelboim, \emph{Quantization of Gauge Systems}, Princeton University Press, New Jersey, 1992.
\bibitem{30}
F. Brandt，\emph{Local BRST Cohomology and Covariance}, Commun. Math. Phys. 1997, 190(2),459–489.
\bibitem{31}
G. Barnich, F. Brandt and M. Henneaux, \emph{Local BRST cohomology in the antifield formalism.I. General theorems}, Commun. Math. Phys. 174 (1995) 57.
\bibitem{32}
G. Barnich, F. Brandt and M. Henneaux, \emph{Local BRST cohomology in the antifield formalism. II. Application to Yang-Mills theory}, Commun. Math. Phys. 174 (1995) 93.
\bibitem{33}
G. Barnich and N. Boulanger, \emph{A note on local BRST cohomology of Yang-Mills type theories with free abelian factors}, J. Math. Phys. 59 (2018) 052302.

\bibitem{34}
I.A. Batalin and G.A. Vilkovisky, \emph{Gauge algebra and quantization}, Phys. Lett. B 102 (1981) 27-31.
\bibitem{35}
I.A. Batalin and G.A. Vilkovisky, \emph{Quantization of gauge theories with linearly dependent generators}, Phys. Rev. D 28 (1983) 2567-2582.
\bibitem{36}
I.A. Batalin and G.A. Vilkovisky,\emph{Existence Theorem for Gauge Algebra},J. Math. Phys. 26 (1985) 172-184.
\bibitem{37}
J. Gomis, J. Par\'{i}s and S. Samuel,\emph{ Antibracket, antifields and gauge theory quantization}, Phys. Rept. 259 (1995).


\bibitem{38}
G. Barnich and M. Henneaux, \emph{Consistent couplings between fields with a gauge freedom and deformations of the master equation}, Phys. Lett. B 311 (1993) 123.
\bibitem{39}
 M. Henneaux, \emph{Consistent interactions between gauge fields: The Cohomological approach}, Contemp. Math. 219 (1998) 93.
\bibitem{40}
F. Brandt,  \emph{Deformations of global symmetries in the extended antifield formalism}, J. Math. Phys. 40 (1999) 1023.
\bibitem{41}
C. Bizdadea, \emph{On the cohomological derivation of topological Yang-Mills theory}, EPL.49:123-129,2000.
\bibitem{42}
C. Bizdadea, L. Saliu and S.O. Saliu,\emph{On Chapline-Manton couplings: a cohomological approach},Phys.Scripta. 61 (2000) 307-310.
\bibitem{43}
C. Bizdadea, E.M. Cioroianu, M.T. Miauta, I. Negru and S.O. Saliu, \emph{Lagrangian cohomological couplings among vector fields and matter fields}, Ann. Phys. 10 (2001) 921-934.
\bibitem{44}
C. Bizdadea, C.C. Ciobirca, E.M. Cioroianu, S.O. Saliu and S.C. Sararu, \emph{Four-dimensional couplings among BF and matter theories from BRST cohomology}, Ann. Phys. 12 (2003) 543-571.
\bibitem{45}
C. Bizdadea, E.M. Cioroianu, S.O. Saliu, S.C. Sararu and M. Iordache, \emph{Four-dimensional couplings among BF and massless Rarita-Schwinger theories: a BRST cohomological approach}, Eur. Phys. J.C, 58:123-149,2008.
\bibitem{46}
C. Bizdadea, E.M. Cioroianu, A. Danehkar, M. Iordache, S.O. Saliu and S.C. Sararu, \emph{Consistent interactions of dual linearized gravity in D=5: couplings with a topological BF model}, Eur. Phys. J. C, 63:491-519, 2009.
\bibitem{47}
C. Bizdadea and S.O. Saliu, \emph{Gauge-invariant massive BF models}, Eur. Phys.J.C, 76:65, 2016.
\bibitem{48}
A. Danehkar, \emph{On the Cohomological Derivation of Yang-Mills Theory in the Antifield Formalism},JHEP. Grav.Cosmol.3:368-387,2017.
\bibitem{49}
N.Boulanger, C. Deffayet, S.G. Saenz and L. Traina, \emph{Consistent deformations of free massive field theories in the Stueckelberg formulation}, JHEP. 1807 (2018) 021.

\bibitem{50}
A.V. Proeyen, \emph{Batalin-Vilkovisky Lagrangian Quantisation}, arXiv:9109036.
\bibitem{51}
K. Sundermeyer, \emph{Constrained Dynamics}, Lecture Notes in Physics Vol. 169, Springer, New York, 1982.
\bibitem{52}
S.C. Sararu, \emph{A first-class approach of higher derivative Maxwell–Chern–Simons–Proca model}, Eur.Phys.J.C 75(11)(2014).

\bibitem{53}
A.V. Smilga, \emph{Benign vs malicious ghosts in higher-derivative theories}, Nucl. Phys. B706 (2005) 598-614.
\bibitem{54}
A.V. Smilga, \emph{Supersymmetric field theory with benign ghosts}, J. Phys. A: Math. Theor. 47 (2014) 052001.
\bibitem{55}
M. Pavsic, \emph{Pais-Uhlenbeck oscillator with a benign friction force}, Phys. Rev. D87 (2013) 107502.





\end{thebibliography}
\end{document}